ORIGINAL RESEARCH

# On the origin of surfaces-dependent growth of benzoic acid crystal inferred through the droplet evaporation method

Piotr Cysewski · Maciej Przybyłek ·
Tomasz Miernik · Mirosław Kobierski ·
Dorota Ziółkowska

**Abstract** Crystal growth behavior of benzoic acid crystals on different surfaces was examined. The performed experiments documented the existence of very strong influence introduced by polar surfaces as glass, gelatin, and polyvinyl alcohol (PVA) on the growth of benzoic acid crystals. These surfaces impose strong orientation effect resulting in a dramatic reduction of number of faces seen with x-ray powder diffractions (XPRD). However, scrapping the crystal off the surface leads to a morphology that is similar to the one observed for bulk crystallization. The surfaces of low wettability (paraffin) seem to be useful for preparation of amorphous powders, even for well-crystallizable compounds. The performed quantum chemistry computations characterized energetic contributions to stabilization of morphology related faces. It has been demonstrated, that the dominant face (002) of benzoic acid crystal, growing on polar surfaces, is characterized by the highest densities of intermolecular interaction energies determining the highest cohesive properties among all studied faces. Additionally, the inter-layer interactions, which stand for adhesive properties, are also the strongest in the case of this face. Thus, quantum chemistry computations providing detailed description of energetic contributions can be successfully used for clarification of adhesive and cohesive nature of benzoic acids crystal faces.

## Introduction

The control of crystallization is essential both from scientific and technological perspectives. Formation of biominerals, production of pharmaceuticals, supermolecular chemistry, semiconductors manufacturing, or designing materials for nonlinear optics or biomineralization are some of potential areas of interests [1–3]. Probably one of the most popular methods of crystallization is solution evaporation. Particularly noteworthy is the droplet evaporative crystallization. In this method, small amounts of solution are evaporated on the surface in order to obtain a layer of crystals. Many studies suggested that this method can be useful in crystallinity, morphology, and polymorphism control [4–9]. Recently, there has been a growing interest in crystallization on different surfaces, especially polymer-induced heteronucleation [5, 10–17]. These studies showed that the nature of crystal growth depends on the polarity of the surface on which crystallization is carried out. According to Shahidzadeh-Bonn et al. [18] in the case

P. Cysewski (✉) · M. Przybyłek · T. Miernik
Department of Physical Chemistry, Collegium Medicum of
Bydgoszcz, Nicolaus Copernicus University in Toruń,
Kurpińskiego 5, 85-950 Bydgoszcz, Poland
e-mail: piotr.cysewski@cm.umk.pl

P. Cysewski · D. Ziółkowska
Faculty of Chemical Technology and Engineering, University of
Technology and Life Sciences in Bydgoszcz, ul. Seminaryjna 3,
85-326 Bydgoszcz, Poland

M. Kobierski
Department of Land Reclamation and Agrometeorology, Faculty
of Agriculture and Biotechnology, University of Technology and
Life Sciences in Bydgoszcz, Bernardyńska 6, 85-029 Bydgoszcz,
Poland

of $Na_2SO_4$, crystal growth starts at solution-air interference irrespectively of the surface polarity. On the other hand, studies on the evaporative microwave and metal-assisted crystallization of L-alanine showed that the type of surface significantly affected crystal shape and size [16]. When crystallization proceeds on a surface which plays the role of heteronucleant, crystals grow mainly in a direction that is perpendicular to the surface. Therefore, it is understandable that the number of diffraction spectra peaks recorded for crystals deposited on different surfaces is significantly smaller than in the case of diffraction patterns recorded for crystals obtained through bulk crystallization and calculated for monocrystals [19–21].

It is well known that aromatic carboxylic acids are of potential pharmaceutical interest. They can be used as anti-inflammatory drugs, as for example acetylsalicylic acid, and some of them exhibit antitumor activity [22]. Also benzoic acid, the simplest representative of this class, is used as an antiseptic agent as well as an additive in food and pharmaceutical industry [23]. It is commonly recognized that the crystal form, which depends on the crystallization conditions, is responsible for many properties like solubility, dissolution, moisture uptake (hygroscopicity), chemical stability (shelf life), hydrate/solvate formation, crystal morphology, fusion properties, thermal stability, mechanical properties, and bioavailability [24–26].

Although the crystal structure of benzoic acid is known [27–29], no reports were published documenting droplet evaporative crystallization. The main goal of this study is to examine and explain the effect of surfaces of varying polarities on benzoic acid crystal growth.

## Methods

### Materials

Analytical grade benzoic acid, methanol, gelatin and polyvinyl alcohol (PVA) (molecular weight 72,000) were purchased from POCH (Gliwice, Poland) and used without further purification.

### Coating procedures

Crystallization of benzoic acid was carried out on blank glass microscope slides and also on slides coated with different materials such as gelatin, PVA, and paraffin. Gelatin and PVA coatings were prepared according to the following procedure. First, 0.5 g of polymer was added to 10 ml of water and constantly stirred for 30 min in 70 °C. Then, 1 ml of prepared mixtures was poured and uniformly spread onto the surface of microscope slides and dried for 24 h under atmospheric pressure at 43 °C. Paraffin layer was prepared by coating microscope slides with Parafilm "M" (American National Can, Greenwich).

### Crystallization

Crystallization via droplet evaporation was performed according to the following procedure. A 20-µL droplet of 0.724 M methanolic solution of benzoic acid was added onto a microscope slide and then evaporated under atmospheric pressure at 43 °C. Bulk crystallization was carried out by evaporating 30 ml of benzoic acid solution at 43 °C in a glass beaker.

### XRD measurements

Powder x-ray diffraction (XRD) was measured using Goniometer PW3050/60 armed with Empyrean XRD tube Cu LFF DK303072. The data were collected in the range between 2° and 40° of $2\theta$ with a 0.001° minimum step size directly on the films and additionally after careful scrapping crystals off the surface. The Reflex program within Accelrys Material studio 6.1 [30] was used for data processing. The background scattering contributions from experimental powder diffraction patterns was calculated and subtracted after $K\alpha_2$ peaks stripping using default settings.

### Computations

The energetic patterns of benzoic acid crystals were analyzed in three steps. Initially, available crystals structures taken from CIF files deposited in CSD [31] were pre-optimized within DMol³ [32–34] module implemented in Accelrys Material Studio 6.1 package [30]. The PBE [35] density functional approach with version 3.5 DNP basis set [36] was used for structure optimization of benzoic acid crystal. This double numerical basis set includes polarization d-function on all non-hydrogen atoms and additionally the p-function on all hydrogen atoms. Such extension is essential for proper hydrogen bonding computations. The dispersion contribution was evaluated based on Grimme [37] approach. The fine option was set for integration accuracy and SCF tolerance ($<10^{-6}$). All electrons were included in core treatment. Also orbital cutoff quality was set to fine. The cell parameters were kept constant at experimental values. This partial optimization enabled the geometry relaxation of benzoic acid molecule without affecting the overall crystal structure. Data characterizing all optimized crystals were collected in supporting materials (see Table S1).

In the second step, the obtained crystals were used for molecular shell preparation and identification of unique pairs formed by two benzoic acid monomers. Taking advantage of the Mercury software [38], the closest proximity of benzoic acid molecule was taken into account for all considered crystals. The nearest neighborhood was defined by separation distance between two monomers, which did not exceed the sum of van der Waals radius augmented by 1 Å of any pair of atoms belonging to either of monomers. This is accepted as a standard procedure [39] defining molecular shell within crystal. In the case of systems with just one molecule per asymmetric unit (Z'), all molecules are supposed to be structurally and energetically identical. The geometries of benzoic acid pairs belonging to molecular shell were used for intermolecular interaction energy (IIE) computations. For this purpose, the *first principle* meta hybrid M06-2X [40] approach along with ET-pVQZ basis set was applied. All values were corrected for basis superposition error. Besides, the interaction energy was decomposed into the following three physically meaningful terms according to Morokuma-Ziegler scheme [41], in which the intermolecular interactions are expressed as the sum of the following bond energy contributions [41], namely

$$\Delta E_{\text{IIE}}^{\text{surf}} = \Delta E_{\text{EL}}^{\text{surf}} + \Delta E_{\text{TPR}}^{\text{surf}} + \Delta E_{\text{OI}}^{\text{surf}}$$

The first term $(\Delta E_{\text{EL}}^{\text{surf}})$ stands for the classical electrostatic interaction between the unperturbed charge distributions of the prepared fragments as they are brought together to their final positions. The second term $(\Delta E_{\text{TPR}}^{\text{surf}})$ stands for the total Pauli repulsion and it accounts for the repulsive interactions between occupied orbitals originating from the Pauli principle and explicit antisymmetrization of the wave function. The last term $(\Delta E_{\text{OI}}^{\text{surf}})$ accounts for stabilization of orbital interactions of one fragment with the unoccupied molecular orbitals of the other one. It also includes mixing of occupied and virtual orbitals within the same fragment resulting in inner-fragment polarization. The decomposition was performed using ADF2013 software [42]. The superscript *surf* resembles the fact that all energy values were divided by surface area of corresponding face. This value was computed in the third step for each morphologically relevant crystal face.

The Miller indices of the most dominant faces were identified based on signals appearing in XPRD spectra. The crystal was cleaved along each of these planes and molecules within $3 \times 3 \times 1$ cell were used as building blocks for surface construction. This surface exposed cell (SEC) was used for large supercell construction by replicating its images along both 2D directions and additionally along the in-depth axis. Practically, the $9 \times 9 \times 9$ supercell system was sufficiently large for edge artifacts elimination and mimicking an infinitely extended surface. All technical manipulations with crystals were performed with an aid of quite convenient facilities offered by Accelrys Material

Studio 6.1 package [30]. The intermolecular interactions between molecules found in SEC, with its images along 2D directions, were used for quantification of intra-layer stabilization and defined the cohesiveness of a particular surface. On the other hand, the inter-layer stabilization, defined by interactions of molecules found in SEC with its images along in-depth direction, was used as a measure of adhesive properties of particular crystal face. Both cohesive and adhesive intermolecular interactions of each considered face were expressed as surface densities estimated by adding up all contributions and dividing them by SEC surface area.

## Results and discussion

### Surface crystallization

Surface evaporative crystallization leads to different morphology than in the case of bulk evaporative crystallization. Indeed, exemplary microscopic images of benzoic acid crystals formed through crystallization on different surfaces that are presented in Fig. 1 confirms this notion. In the case of polar surfaces, typical fern-like patterns are developed with varying textures and intensities. Individual crystals formed on non-polar surfaces, due to fast crystal growth, are much smaller than the ones formed via bulk crystallization.

The quantitative analysis of obtained crystals was conducted based on measured PXRD spectra. The collected data were plotted in Fig. 2. First of all, the pattern characterizing bulk crystallization of benzoic acid agrees with those documented in the literature [27–29]. Apart from bulk crystallization spectrum, there are also provided spectra corresponding directly to the surface crystallization and additionally to the same crystals measured after carefully scrapping them off the surface. This was done for checking if there is any orientation effect imposed by specific arrangement of crystals on the surface. As it was documented in Fig. 2, the influence of surface on the morphology of obtained crystals appears to be extremely strong. In the studied range of diffraction angles, the benzoic acid crystals obtained on different surfaces show only few strong reflections. In all cases, the most intense band is located at 8.1° corresponding to Miller plane (002) with d-spacing distance $d_{002} = 10.9$ nm. There are also visible two less intense peaks related to (100) and (10-2) faces associated with $2\theta = 16.3°$ and $17.3°$, respectively. The rest of signals typical for fully developed benzoic acid crystal morphology are not visible. This is especially spectacular for highly polar surfaces as glass or PVA. Interestingly, the powder obtained after scrapping crystals

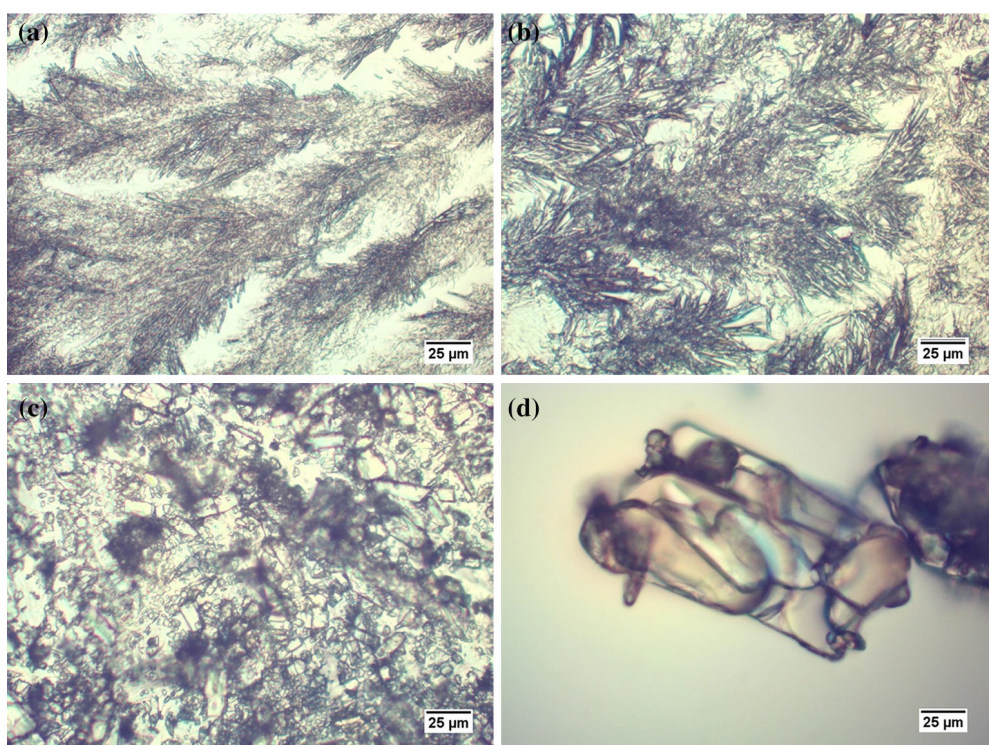

**Fig. 1** The representative microscope images of the benzoic acid crystals formed through crystallization on different surfaces, namely glass (**a**), PVA (**b**), paraffin (**c**), and after bulk crystallization (**d**)

off the polar surface retains bulk-like patterns. This suggests that the observed reduction of signals is possibly mainly due to the crystals orientations effect. Due to the interactions of benzoic acid molecules with the polar surface, only certain crystal faces can grow perpendicularly to the surface and hence the majority of faces are not available for x-ray irradiation scattering. This is even more visible for PVA surfaces, since after mechanical removal from the surface, the obtained precipitate has its morphology very similar to the one obtained from bulk crystallization. The way of mechanical treatment of crystals during scrapping off is probably important for final morphology profile. For gelatin coatings, a similar effect is observed, but in general crystallization is much less effective and the intensities of measured PXRD signals are much lower compared to the other surfaces. It seems that on gelatin surface in oversaturated conditions the crystal formation of benzoic acid is prohibited and mostly amorphous phase is obtained. This is an interesting observation since amorphous phases are important from a pharmaceutical perspective. The PXRD spectrum recorded for crystals produced via evaporative crystallization on paraffin seems to be more similar to diffraction patterns recorded for crystals obtained through bulk evaporative crystallization. Because of the low wettability of paraffin, benzoic crystal faces growth proceeds in all directions without directional effect, in contrast to crystallization on polar surfaces.

## Origin of the crystal growth and orientation effect

Before explaining the growth of benzoic acid crystals on the surface, the accuracy of computations was carefully checked against experimental data of sublimation enthalpy. Benzoic acid structures measured at ambient conditions are deposited in CSD three times under BENZAC, BEZAC01, and BENZAC02 [27–29] codes. Although all of them correspond to monoclinic crystals system, characterized by P21/c space group, they differ in many details. For example, the cell volume ranges from 616.733 $Å^3$, through 619.15 $Å^3$, to 613.955 $Å^3$, respectively. This is not only associated with variation of cell parameters but also with changes of all crucial features affecting intermolecular interactions. For example, the main $C_2^2(8)$ synthon geometry is significantly different in these structures since hydrogen bond length is equal to 2.616, 2.633, and 2.627 Å for these three mentioned structures, respectively. That is why the geometries provided by CIF files seem to be inadequate for direct energy estimations. Optimization of molecular structures without changing cell parameters leads to a much better congruency of molecular geometry (2.601, 2.608 and 2.603 Å, respectively). Additionally, the hydrogen atom position within $C_2^2(8)$ synthon of benzoic acid derivatives is questionable [43] due to concurrency between electronegative centers, leading to hydrogen atoms disorder [27–29, 44–46]. Fortunately, these two

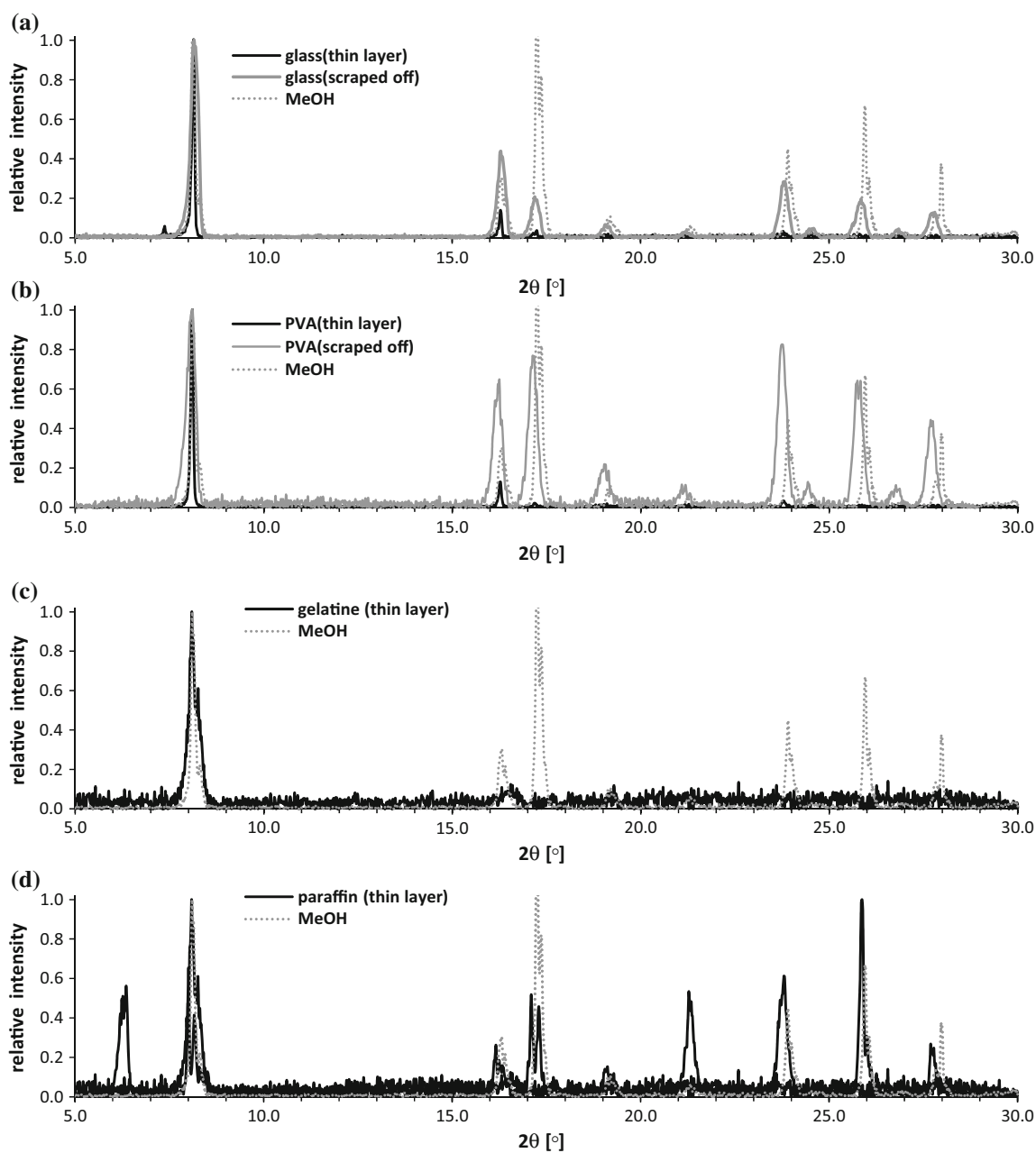

**Fig. 2** The powder x-ray spectra of benzoic acid thin layers formed on **a** glass, **b** polyvinyl alcohol (PVA), **c** gelatin, and **d** paraffin. MeOH denotes bulk evaporative crystallization

tautomeric forms are energetically almost identical and their XPRD spectra are indistinguishable. However, all six optimized structures were used for averaging of intermolecular interactions characterizing distinct contacts within molecular shell. The structural characteristics of optimized crystals are provided in supporting materials in Table S1. In case of benzoic acid crystal, each molecule is surrounded by 15 neighbors belonging to the molecular shell. However, only 9 intermolecular contacts are distinct and univocally define all energetic patterns within both bulk crystal and morphology related faces. The values of intermolecular interactions of these contacts are provided in supporting materials in Table S2. These data can be used for additive construction of stabilization energies of the whole crystal by simple summing of unique pair interaction energies ($\varepsilon_{HE}$) weighted by their occurrence ($n_{ij}$), namely $E_{latt} \approx \Delta E_{MS} = 0.5 \bullet \Sigma n_{ij} \bullet \varepsilon_{HE}(ij)$ (0.5 factor is used in order to avoid double counting of intermolecular interactions in crystal). For validation of reliability of the additive model, the experimental values of sublimation enthalpies

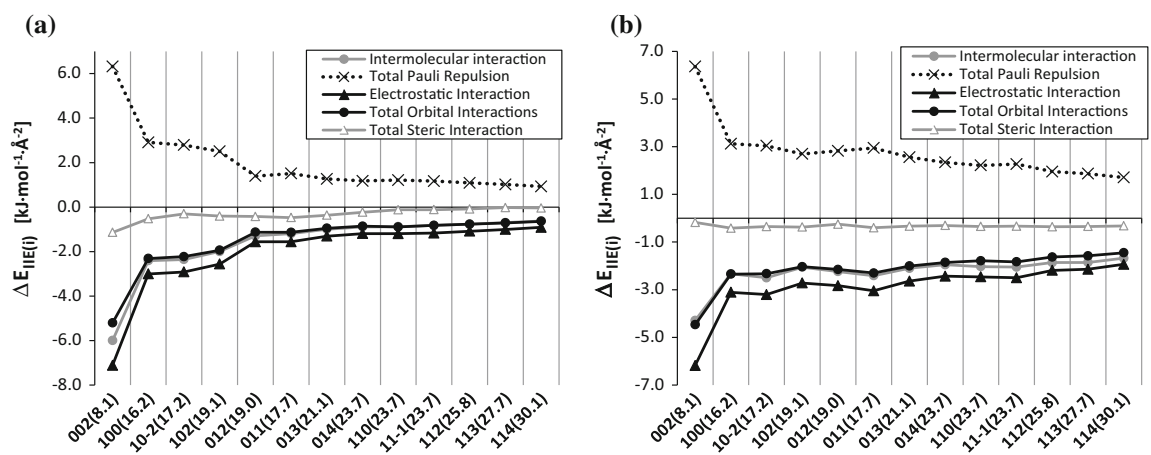

**Fig. 3** Surface densities of **a** cohesive and **b** adhesive intermolecular interactions of morphology dominant faces of benzoic acid crystal. Miller indices along with $2\theta$ values (in parenthesis) were provided in the description of abscissa

were used. Since the sublimation enthalpy, $\Delta H_{sub}(T)$, of a crystal is a direct measure of the lattice energy, these data are very often used for theoretical models verification [47, 48] by the following simplified formula: $\Delta H_{sub}(T) = -E_{latt} - 2RT$, where $T$ is the temperature at which the sublimation enthalpy is measured and $R$ stands for the gas constant. This equation relies on the assumption, that the gas phase is ideal and energy contributions from intramolecular vibrational motions are equal in the solid and in the gas phases. It is worth mentioning, that there are systematic uncertainties associated with the measurement of sublimation enthalpies of solids and it is common to notice quite large variances in sublimation enthalpy of the same compound [49]. Benzoic acid is used as internal standard in thermo-gravimetric measurements and for this purpose precise values of heats of phase changes are indispensable. This is the reason why measurements were repeated by different authors using a variety of experimental approaches. As reported in contemporary compilation [49] and Reaxys database, there were published 47 measurements of benzoic acid sublimation enthalpies. After adjustments to standard temperature, the average value is equal to 89.3 kJ/mol with averaged standard deviation equal to 2.3 kJ/mol [49]. The corresponding value of lattice energy equals $-94.3 \pm 2.3$ kJ/mol and is used here for reference purposes. Interestingly, the additive model used for benzoic acid crystals characteristics is sufficiently accurate since it predicts the lattice energy as equal to $-95.9 \pm 0.3$ kJ/mol.

Experimentally observed directional effect of polar surfaces on orientations of benzoic acid crystals is an intriguing aspect, which deserves further exploration. For closer inspection into the origin of such behavior, the model surfaces were constructed according to procedure described in the methodology part. The face-related profiles of pair interactions were generated and energy components were computed. Results of these computations are collected in Fig. 3. The most interesting outcome of performed computations is the identification of the unique features of the (002) face. First of all, the surface density of intra-intermolecular interaction is the highest for this particular crystal face. Thus, this crystal face is characterized by the strongest cohesive forces stabilizing such layers, what is quite understandable since the dominant contribution to the energy comes from $C_2^2(8)$ synthons formed by each benzoic acid dimers. Interestingly, the surface densities of the adhesive interactions are also the most attractive for (002) face as it can be seen on the right panel of Fig. 3. Again, the exposure of carboxylic groups outwards is the source of this property. Thus, the unique features of (002) face of benzoic acid crystals are related to the highest interaction densities both of cohesive and adhesive character, which means that the most dominant crystal growth direction is also the most energetically favorable. This is why this particular face overwhelmingly dominates in the case of crystallization on polar surfaces. The highest density of IIE characterizing the (002) face is also associated with the highest electrostatic and orbital interactions contributions. Although the total Pauli repulsion is also the highest for this face, electrostatic interactions are dominant and overcome the steric repulsions.

The energy decomposition offered by Morokuma–Ziegler scheme leads to quantities strongly related to each other. As one can see in Fig. 3, the trends of electrostatic, total orbital, and intermolecular interactions contributions are arranged almost symmetrically with respect to total Pauli repulsion contribution. In Fig. 4, the correlation between the values of the total Pauli repulsion, $\Delta E_{TPR}$ and some selected contributions are presented, showing highly linear correlations. First of all, the rise of total Pauli

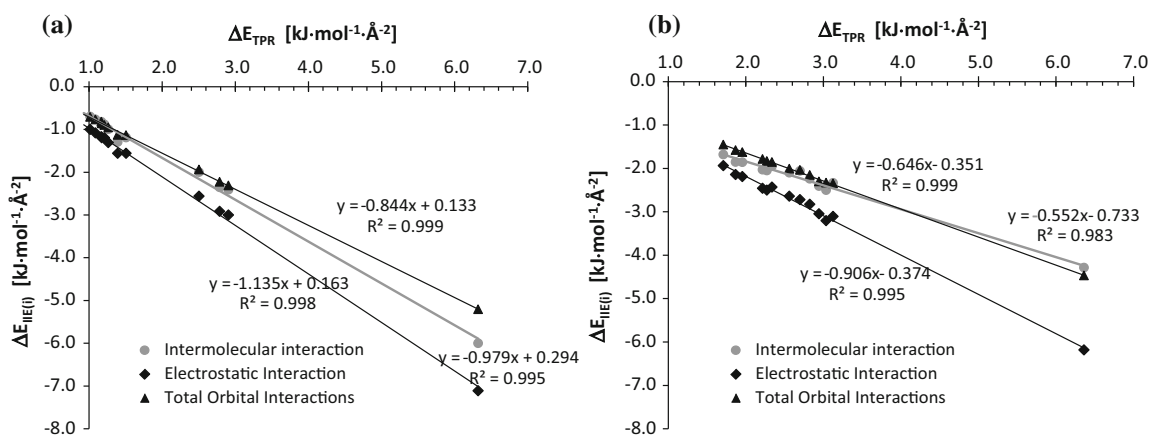

**Fig. 4** Correlations between contributions to cohesive (**a**) and adhesive (**b**) interactions of morphology dominant faces of benzoic acid crystal

repulsion in the diverse molecular conformations of analyzed dimers is associated with an increase of all electrostatic and orbital attractions and consequently an increase of pair stabilization. This is quite expected, since stronger intermolecular interactions require more robust alterations of electronic structure of interacting monomers. The nature of Pauli repulsion interactions, also known as exchange interactions, is associated with the repulsion of electrons with the same spin. According to Lenard-Jones model, both repulsion and attraction energies depend on the intermolecular distances. Therefore, increase of attraction interactions contributes to the shortening of distances between molecules and in turn an increase of steric repulsion. Interestingly, the relationship between Pauli repulsion contribution and other types of IIE densities is not identical for cohesive and adhesive forces. This can be inferred from Fig. 4 by inspection of slope values of presented linear regressions. The only value greater then unity has been found for electrostatic contributions to cohesive forces. This suggests that in the case of intra-layer contacts the rise of the Pauli repulsions is associated with even more pronounced electrostatic attractions. However, this is not observed in the case of adhesive interactions. In general, slopes of trends presented in Fig. 4 for cohesive interactions are almost twice as high as adhesive ones. Thus, intermolecular interactions are much more sensitive to Pauli repulsions in the case of inter-layer interactions and this cannot be merely related to the range of $\Delta E_{TPR}$ values. The span of densities of total Pauli repulsion is very similar for both cohesion and adhesion interactions.

## Conclusions

Benzoic acid stands here as a model compound constituting a class of aromatic carboxylic acids. Since many drugs belong to this group, studying the properties of benzoic acid is a valuable starting point. It is quite rational to expect that such active pharmaceutical ingredients as aspirin, salicylic acid, diflunisal, flufenamic acid, lasalocid, and others can also behave on similar manner. Controlling morphology is an important aspect of crystal engineering. Among many possible ways of achieving such control the crystallization on surfaces through droplet evaporation can be applied. The performed experiments on crystallization of benzoic acid suggested the existence of very strong effect on crystal growth behavior and resulting apparent morphology, especially introduced by polar surfaces. This suggests the possibility of controlling crystallization in two aspects. First of all, the polar surfaces impose strong orientation effect associated with a dramatic reduction of number of faces seen with x-ray diffractions. However, scrapping the crystal off the surface leads to reestablishing a morphology similar to the one observed for bulk crystallization. Despite the fact, that actual morphology of the surface can be quite similar as after evaporative crystallization, the orientation effect changes physical properties of such deposits. As long as they adhere to polar surface, they have the properties determined by a dominant face exposed to solution. Second interesting aspect was inferred form crystallization on gelatin and especially on paraffin. These surfaces prohibited normal crystal growth leading mostly to amorphous phases. Solid deposits obtained in such way are not rich in crystals and observed XPRD intensities are at least one order of magnitude lower compared to crystallization on glass surface. This observation can offer practical applications. Those surfaces of low wettability like paraffin can be used for preparation of powder, even from well crystallizable compounds, with high contributions from amorphous phases. In order to reveal the mechanism governing the above observations, there were performed computations characterizing the energetic contributions stabilizing particular crystal faces. Quite simple and clear picture emerged from these computations, which is also in good accordance with chemical intuition. Indeed, the dominant faces observed on polar surfaces and emerging due to orientation effect are characterized by the highest densities of intermolecular interaction energies, which determine cohesive properties. Additionally, the inter-layer interactions, which stand for adhesive properties, are also the strongest in case of (002) face of benzoic acid crystal. Thus, quantum chemistry computations, providing detailed description of energetic contributions, can be successfully used for clarification of adhesive and cohesive nature of benzoic acid crystal faces.

**Acknowledgments** This research was supported in part by PL-Grid Infrastructure. The allocation of computational facilities of Academic Computer Centre "Cyfronet" AGH/Krakow/Poland is also acknowledged.

## References

1. Debenedetti PG (1996) Metastable liquids: concepts and principles. Princeton University Press, Princeton
2. Mullin JW (2001) Crystallization, 4th edn. Butterworth-Heinemann, Oxford, Boston
3. Katzsch F, Eißmann D, Weber E (2012) Struct Chem 23:245–255
4. Carver KM, Snyder RC (2012) Ind Eng Chem Res 51:15720–15728
5. Araya-Sibaja AMA, Fandaruff C, Campos CEM, Soldi V, Cardoso SG, Cuffini SL (2013) Scanning 35:213–221
6. Chiou D, Langrish TAG, Braham R (2008) J Food Eng 86:288–293
7. Islam MIU, Langrish TAG (2009) Food Bioprod Process 87:87–95
8. Langrish TAG (2009) J Food Eng 91:521–525
9. Beckmann W, Otto WH (1996) Trans Inst Chem Eng 74:750–758
10. Mojibola A, Dongmo-Momo G, Mohammed M, Aslan K (2014) Cryst Growth Des 14:2494–2501
11. Grzesiak AL, Matzger AJ (2008) Cryst Growth Des 8:347–350
12. Curcio E, López-Mejías V, Di Profio G, Fontananova E, Drioli E, Trout BL, Myerson AS (2014) Cryst Growth Des 14:678–686
13. Lupi L, Hudait A, Molinero V (2014) J Am Chem Soc 136:3156–3164
14. Diao Y, Myerson AS, Hatton TA, Trout BL (2011) Langmuir 27:5324–5334
15. López-Mejías V, Kampf JW, Matzger AJ (2009) J Am Chem Soc 131:4554–4555
16. Alabanza AM, Pozharski E, Aslan K (2012) Cryst Growth Des 12:346–353
17. Grell TAJ, Pinard MA, Pettis D, Aslan K (2012) Nano Biomed Eng 4:125–131
18. Shahidzadeh-Bonn N, Rafaï S, Bonn D, Wegdam G (2008) Langmuir 24:8599–8605
19. López-Mejías V, Knight JL, Brooks CL, Matzger AJ (2011) Langmuir 27:7575–7579
20. Curcio E, López-Mejías V, Di Profio G, Fontananova E, Drioli E, Trout BL, Myerson AS (2014) Cryst Growth Des 14:678–686
21. Diao Y, Myerson AS, Hatton TA, Trout BL (2011) Langmuir 27:5324–5334
22. Djurendić EA, Klisurić OR, Szécsi M, Sakač MN, Jovanović-Šanta SS, Ignáth I, Kojić VV, Oklješa AM, Savić MP, Penov-Gaši KM (2014) X-ray structural analysis and antitumor activity of new salicylic acid derivatives. Struct Chem. doi:10.1007/s11224-014-0450-2
23. Cherrington CA, Hinton M, Mead GC, Chopra I (1991) Adv Microb Physiol 32:87–108
24. Krishna EH, Gupta VRM, Samreen NS, Jyothi S (2013) Der Pharm Sin 4:77–87
25. Rodrígue -Hornedo N, Nehm SJ, Jayasankar A (2006) In: Swarbrick (ed) Encyclopedia of pharmaceutical technology, 3rd edn. Informa Health Care, New York
26. Rodríguez-Hornedo N (2007) Mol Pharmaceutics 4:299–300
27. Sim GA, Robertson JM, Goodwin TH (1955) Acta Crystallogr 8:157–164
28. Bruno G, Randaccio L (1980) Acta Crystallogr B 36:1711–1712
29. Feld R, Lehmann MS, Muir KW, Speakman JC (1981) Z Krist 157:215–231
30. Accelrys (2013) Materials studio 6.1. Accelrys, San Diego
31. CSD Version 5.32. Updated May 2011
32. Delley B (2000) J Chem Phys 113:7756–7764
33. Delley B (1996) J Phys Chem 100:6107–6110
34. Delley B (1990) J Chem Phys 92:508–517
35. Perdew JP, Burke K, Ernzerhof M (1996) Phys Rev Lett 77:3865–3868
36. Delley B (2006) J Phys Chem 110:13632–13639
37. Grimme S (2006) J Comput Chem 27:1787–1799
38. Macrae CF, Edgington PR, McCabe P, Pidcock E, Shields GP, Taylor R, Towler M, van de Streek J (2006) J Appl Cryst 39:453–457
39. Shishkin OV, Dyakonenko VV, Maleev AV (2012) CrystEngComm 14:1795–1804
40. Zhao Y, Truhlar DG (2008) Theor Chem Acc 120:215–241
41. Bickelhaupt FM, Baerends EJ (2000) In: Lipkowitz KB, Boyd DB (eds) Rev Comput Chem, vol 15. Wiley-VCH, New York
42. ADF (2013) SCM. Vrije Universiteit, Amsterdam, The Netherlands
43. Wilson CC, Shankland N, Florence AJ (1996) Chem Phys Lett 253:103–107
44. Kanters JA, Roclofsen G, Kroon J (1975) Nature 257:625–626
45. Fischer P, Zolliker P, Meier BH, Ernst RR, Hewat AW, Jorgensen JD, Rotella FJ (1986) J Solid State Chem 61:109–125
46. Dcstro R (1991) Chem Phys Lett 181:232–236
47. Gavezzotti A, Filippini G (1997) In: Gavezzotti A (ed) Theoretical Aspects and Computer Modeling. Wiley, New York
48. Gavezzotti A (1994) In: Burgiand H-B, Dunitz D (eds) Structure Correlation, vol 2. VCH, Weinheim, pp 509–542
49. Acree W Jr, Chickos JS (2010) J Phys Chem Ref Data 39:043101-1–043101-942